\documentclass[11pt]{article}

\usepackage{times}
\usepackage{a4wide}
\usepackage[final]{graphicx}
\usepackage{amsmath,amssymb}
\usepackage{authblk} 
\usepackage[numbers,compress]{natbib}
\usepackage[margin=1em,font={small},labelfont={sc}]{caption}

  \renewenvironment{thebibliography}[1]{%
    \begin{oldthebibliography}{#1}%
      \setlength{\parskip}{0ex}%
      \setlength{\itemsep}{0ex}%
      \small
  }%
  {%
    \end{oldthebibliography}%
  }

\newcommand{\Q}{{\cal Q}}
\newcommand{\xB}{x_{\rm B}}
\newcommand{\GeV}{{\rm GeV}}
\newcommand{\cffF}{{\cal F}}
\newcommand{\cffH}{{\cal H}}
\newcommand{\cffE}{{\cal E}}
\newcommand{\cfftH}{\widetilde{\cal H}}

\newcommand{\cffbE}{\overline{\cal E}}

\newcommand{\Asin}[2]{A^{\sin(#1\phi)}_{\rm #2}}
\newcommand{\Acos}[2]{A^{\cos(#1\phi)}_{\rm #2}}
\newcommand{\Asinsin}[2]{A^{\sin(\varphi)\sin(#1\phi)}_{\rm #2}}
\newcommand{\Asincos}[2]{A^{\sin(\varphi)\cos(#1\phi)}_{\rm #2}}
\newcommand{\Acossin}[2]{A^{\cos(\varphi)\sin(#1\phi)}_{\rm #2}}
\newcommand{\Acoscos}[2]{A^{\cos(\varphi)\cos(#1\phi)}_{\rm #2}}

\begin{document}

\title{\LARGE \bf Revealing CFFs and GPDs from experimental measurements%
\footnote{Talk given by D.M. at the {\em 3rd Workshop on the QCD Structure of the Nucleon},
22-26  October 2012, Bilbao, Spain.}%
}


\author[1]{K.~Kumeri\v{c}ki}
\author[2]{D.~M\"{u}ller}
\author[3]{M.~Murray}

\affil[1]{Department of Physics, University of Zagreb - Zagreb, Croatia} 
\affil[2]{Institut f\"ur Theoretische Physik II, Ruhr-Universit\"at Bochum - Bochum, Germany} 
\affil[3]{School of Physics and Astronomy, University of Glasgow - Glasgow, Scotland, UK} 

\date{}

\maketitle

\begin{abstract}
\noindent
We report on the status of the phenomenological access of generalized parton distributions  from photon and meson electroproduction off proton.
Thereby, we emphasize the role of HERMES data for deeply virtual Compton scattering, which allows us to map various asymmetries into the space of Compton form factors.
\end{abstract}

\section{Introduction}
Motivated by the aim of understanding the decomposition of the nucleon spin and resolving the transverse distribution of partons,
large experimental effort has been expended to
measure various observables in the exclusive electroproduction of photons and mesons at medium and high center-of-mass energies,
which have taken place at H{\sc era} and Jefferson Lab. Thereby, deeply virtual Compton scattering process (DVCS)
is viewed as a golden channel, allowing a clean access to generalized parton distributions (GPDs). Besides the DVCS process
the Bethe-Heitler (BH) bremstrahlungs process has the same initial and final states as DVCS ($ep\rightarrow ep\gamma$).
Since the BH amplitude is exactly known to leading order accuracy in the electromagnetic fine structure constant $\alpha_{\rm em}$, it may serve as a reference for the DVCS amplitude. At fixed target kinematics  one utilizes the fact that the large BH contribution in the interference term amplifies the contribution from the more interesting DVCS process. This gives access to linear combinations of Compton form factors (CFFs), allowing to extract both their modulus and the phase.
In collider kinematics the DVCS amplitude overwhelms the BH one, however, also here one may access the interference term.

On the theoretical side the access to GPDs from deeply virtual meson production (DVMP) and DVCS, i.e.,
\begin{eqnarray}
\gamma^\ast_L(q_1)\,p(p_1,s_1) \to N(p_2)\,M(q_2)\;\;  \mbox{and}\;\; \gamma^\ast(q_1)\, p(p_1,s_1) \to p(p_2)\, \gamma(q_2),
\end{eqnarray}
measurements relies on factorization theorems \cite{Collins:1996fb,Collins:1998be}, which are perturbatively
proven to leading order accuracy in $1/\Q^2$. These theorems state that the longitudinal helicity amplitude for DVMP
(transverse helicity amplitude for DVCS) factorizes in GPDs and meson distribution amplitudes (DAs) (final photon state in DVCS has a point-like coupling),
which are process-independent non-perturbative functions,
and a hard scattering amplitude. They also state that non-factorizable final state interaction is suppressed by (at least) an additional power $1/\Q$.
Furthermore, the hard amplitude can be systematically calculated as expansion w.r.t.~QCD coupling  $\alpha_s$,
where the process-independent collinear singularities are factorized out and dress the bare GPDs and eventually also DAs. The theoretical framework
for the processes of interest has been set up for some time to next-to-leading order (NLO) accuracy%
\footnote{If one describes only DVCS,
no essential improvement will be reached by going from LO to NLO, since this can be
absorbed by redefinition of convention-dependent GPDs.
Contrarily, in a global analysis of both DVMP and DVCS it is important to utilize the NLO framework.}, see references in \cite{Belitsky:2005qn}.

The present phenomenological challenge is to describe these exclusive measurements in terms of GPDs.
In Sect.~\ref{sec:observables} we introduce the cross sections in terms of transition and Compton form factors and  we shortly
report on the status of the phenomenology. In
Sect.~\ref{sec:HERMES} we consider  the extraction of  CFFs at
given kinematical points from the H{\sc ermes} measurements as a map of random variables and from the regression approach
and use the H{\sc ermes} data to access CFFs by least squares fitting.  We also present a global GPD model fit
that additionally includes H{\sc era} collider and Jefferson Lab measurements. Finally, we summarize.

\section{GPDs from helicity dependent transition and Compton form factors }
\label{sec:observables}

In DVMP only the (polarized) longitudinal photoproduction cross section
\begin{eqnarray}
\qquad \frac{d\sigma^{\gamma_{\rm L}^*\,p\to M\,N}}{dt d\varphi} =
  \frac{2\pi \alpha_{\rm em}}{\Q^4 \sqrt{1+ \frac{4 \xB^2 M_N^2}{\Q^2}}}\,\frac{\xB^2}{1 - \xB}
  \Big\{{\cal C}_{\rm unp}({\cal F}_M,{\cal F}_M^\ast) + \Lambda \sin(\varphi)\, {\cal C}_{\rm TP}({\cal F}_M,{\cal F}_M^\ast)\Big\},
\label{dX-DVMP}
\end{eqnarray}
for a transversally polarized proton allows to measure longitudinal helicity transition form factors (TFFs) ${\cal F}_M$ that are systematically factorizable in
GPDs and meson DAs.
Here $\Lambda$ is the polarizability of the  polarized proton, $\varphi$ describes the direction of the transverse polarization vector, $\xB$ is the Bjorken
variable, and $\Q^2=-q_1^2$. In these processes the produced meson $M$
serves as a flavor and parity filter. We may define  parity even TFFs (e.g., longitudinally polarized vector mesons $\rho,\omega,\phi$) and parity odd TFFs (e.g., pseudo scalar mesons $\pi,\eta$)  in terms of Dirac bilinears:
\begin{eqnarray}
\label{tffF-def}
\epsilon_L^\mu \langle M N| j_\mu |N\rangle = \left\{\!{
\overline{u}(p_2,s_2) \bigg[
 \frac{{\not\, q}}{P\cdot q}{\cal H}_{M} +  \frac{i \sigma_{\alpha \beta}\, q^\alpha \Delta^\beta}{2 P\cdot q\, M_N}{\cal E}_{M} \bigg]u(p_1,s_1)
\;\;\mbox{(even parity)}
 \atop
 \overline{u}(p_2,s_2) \bigg[
 \frac{{\not\,q}\, \gamma_5}{P\cdot q} \widetilde{\cal H}_{M} +  \frac{\gamma_5\, q\cdot\Delta}{2 P\cdot q\, M_N} \widetilde{\cal E}_{M} \bigg]u(p_1,s_1)
 \;\;\mbox{(odd parity)}
 }\right. ,
\end{eqnarray}
where $\Delta^\mu= p_2^\mu-p_1^\mu = q_1^\mu-q_2^\mu $ is the momentum transfer in the $t$-channel ($t\equiv \Delta^2$) and
$q^\mu/P\cdot q = (q_1^\mu +q_2^\mu)/(q_1+q_2)\cdot(p_1+p_2)$ is a crossing-symmetric auxiliary vector. 
The unpolarized part in (\ref{dX-DVMP}) depends on the squared moduli
$|{\cal H}_M - \xB^2\cdots {\cal E}_M|^2$ and  $|{\cal E}_M|^2$ (same for parity odd case), while the transverse part is proportional to
$\Im{\rm m} {\cal H}_M {\cal E}^\ast_M$ (or $\Im{\rm m} \widetilde{\cal H}_M \widetilde{\cal E}^\ast_M$), i.e., to the  phase difference
of ${\cal H}_M$ and ${\cal E}_M$. Based on the $t$-channel exchange picture, various models have been proposed and are
utilized to describe DVMP processes.

In DVCS only the GPDs enter and  one can access in principle both the modulus and phase of all twelve CFFs (or helicity amplitudes).
However, the extraction of these information requires a complete measurement of cross sections or
asymmetries with all possible polarization options.
Thereby, the fivefold electroproduction cross section,
\begin{eqnarray}
\label{dX^eN2eNgamma}
\qquad
\frac{d^5\sigma}{d\xB d\Q^2 dt d\phi d\varphi } = \frac{\alpha_{\rm em}^3 \xB y^2}{16\pi^2 \Q^4 \sqrt{1+ \frac{4\xB^2 M_p^2}{\Q^2}}} \left[
\frac{|{\cal T}_{\rm BH}|^2}{e^6} \pm \frac{{\cal I}(\cffF)}{e^6} + \frac{|{\cal T}_{\rm DVCS}|^2(\cffF^\ast,\cffF)}{e^6}
\right],
\end{eqnarray}
consists of the BH squared term, the interference term $\cal I$ (linear in CFFs), which is charge odd,
and the DVCS squared term (bilinear form of CFFs), where $y$ is the fractional electron energy loss and $\phi$ an azimuthal angle. The functional
form of both the interference term and DVCS amplitude squared is  known  as function of twelve complex valued helicity dependent CFFs ${\cal F}_{++}$, ${\cal F}_{0+}$, and ${\cal F}_{-+}$,  where
${\cal F}\in \{{\cal H},{\cal E},\widetilde{\cal H},\widetilde{\cal E}\}$ and
subscripts label the helicities of initial ($+,0,-$) and final ($+,-$) state photons \cite{Belitsky:2012ch}.
To LO accuracy the twist-two associated CFFs are given by the charge even quark GPDs
\begin{equation}
\label{LO}
{\cal F}_{++}\approx {\cal F} \stackrel{\rm LO}{=} \sum_{q=u,d,s,\cdots} \int_{-1}^1\!dx\,\frac{e_q^2}{\xi-x -i\epsilon} F^{q^{(+)}}\,,
\quad  \xi\simeq \frac{\xB}{2-\xB}\,,
\end{equation}
where $e_q$ are the fractional quark charges.

DVCS data for unpolarized proton target has been analyzed in global fits \cite{Kumericki:2009uq}.
In particular in the small-$\xB$ region flexible GPD models are needed  and are used
to control both the size and the evolution flow of Compton form factors (CFFs).
Thereby, sea quark and gluon GPDs were directly parameterized in terms of (conformal) GPD moments
rather than in momentum fraction representation. For the analyzes of fixed target measurements the $\Q^2$ evolution
can be neglected. Thus, instead of the LO convolution formulae (\ref{LO}) we can equivalently employ the dispersion relations where
one can directly model the imaginary part of valence GPDs on the cross-over line as function of $\xB$ and $t$ and
possible subtraction constants as function of $t$.
Apart from some earlier model dependent estimates as well as more recent data  descriptions for $\pi^+$ \cite{Bechler:2009me} and light vector mesons \cite{Meskauskas:2011aa} at LO accuracy, the collinear framework has still not been confronted with the increasing amount of experimental
DVMP data. We would like to emphasize that a GPD inspired  hand-bag model approach (or two parton $t$-channel exchange picture)
has been used to confront GPD models with DVMP measurements \cite{Goloskokov:2005sd,Goloskokov:2007nt,Goloskokov:2009ia}. Here,
GPDs are based on the popular Radyushkin ansatz \cite{Radyushkin:1997ki} and NLO parton distribution function
parameterizations with variable $\Q^2$-dependence.
Furthermore, utilizing this model for the dominant GPD $H$ reproduces at LO the collider DVCS data \cite{Meskauskas:2011aa}
and provides the typical predictions for fixed target DVCS data that are known from models based on the Radyushkin ansatz \cite{Kumericki:2011zc}. Very similar results are obtained if one utilizes the complete GPD content of this model for polarized proton DVCS data \cite{Kroll:2012sm}.

\section{CFFs from HERMES measurements and global DVCS fits}
\label{sec:HERMES}

The elementary problem in analyzing DVCS (and also DVMP) data is that the
number of CFFs (TFFs), times two because they are complex quantities, is usually larger than
the number of observables at a given kinematical point.
One must thus rely on model assumptions or hypotheses
which means that, independently of the applied method or
framework, a theoretical bias cannot be avoided in analyzing
the present available world data set.
Fortunately, the DVCS experiments at  H{\sc ermes} had both electron and positron beams available and is currently the
experiment that has delivered the most complete set of DVCS asymmetries in twelve kinematical bins, see Table \ref{tab:means}.
\begin{table}[t]
\begin{center}
\begin{tabular}{|c|c|c|c||c|c|c|c||c|c|c|c|}
  \hline
  1 & 2 & 3 & 4 & 5 & 6 & 7 & 8 & 9 & 10 & 11 & 12 \\
   \hline
 0.03& 0.1\phantom{0}& 0.2\phantom{0}& 0.42& 0.1\phantom{0}& 0.1\phantom{0}& 0.13& 0.2\phantom{0}& 0.08& 0.1\phantom{0}& 0.13& 0.19 \\
 0.08& 0.1\phantom{0}& 0.11& 0.12& 0.05& 0.08& 0.12& 0.2\phantom{0}& 0.06& 0.08& 0.11& 0.17 \\
 1.9\phantom{0}& 2.5\phantom{0}& 2.9\phantom{0}& 3.5\phantom{0}& 1.5\phantom{0}& 2.2\phantom{0}& 3.1\phantom{0}& 5.0\phantom{0}& 1.2\phantom{0}& 1.9\phantom{0}& 2.8\phantom{0}& 4.9\phantom{0} \\
  \hline
\end{tabular}
\end{center}
\vspace{-3mm}
\caption{\label{tab:means}\small
Kinematical mean values for $-t\; [\GeV^2]$ (second row), $\xB$ (third row), and $\Q^2\;  [\GeV^2]$ (forth row) of three times four H{\sc ermes} bins from ref.~\cite{Airapetian:2008aa}, labeled as $\#1,\cdots,\#12$ (first row).
}
\vspace{-5mm}
\end{table}
These data can be locally analyzed by regression methods \cite{Guidal:2010de} or simply mapped into the space of CFFs \cite{Kumericki:2013br}.

Let us explain for a spin-zero target, where we have only three CFFs ${\cal H}_{++}$, ${\cal H}_{0+}$, and ${\cal H}_{-+}$, that asymmetry measurements can be mapped to CFFs, however,  {\em two}
such maps exist.
\begin{figure}[t]
\begin{center}
\includegraphics[width=13cm]{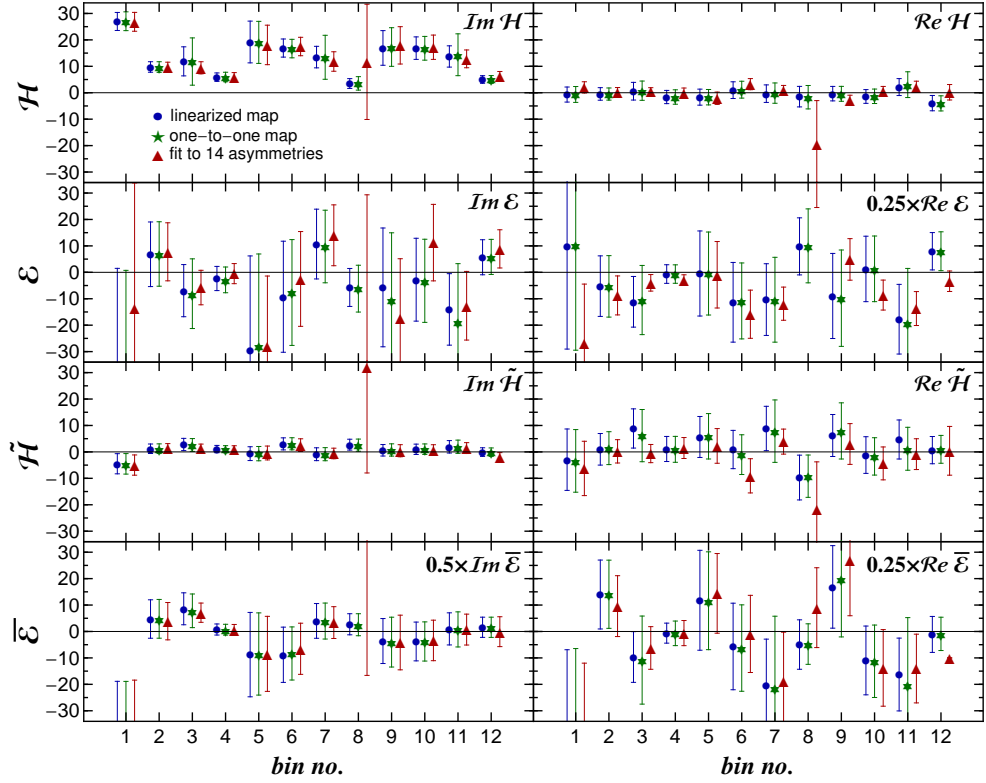}
\end{center}
\vspace{-0.4cm}
\caption{\small CFFs from a linearized (circles, shifted to the left) and a one-to-one map (stars)
of eight twist-two dominated charge odd asymmetries as well as from
a least squares fit (triangles,  shifted to the right)
to fourteen twist-two related observables for each of 12 H{\sc ermes} bins.}
\label{fig:methods}
\vspace{-5mm}
\end{figure}
As for HERMES data we consider the second and third harmonics compatible
with zero, which suggest that the photon helicity flip CFFs, associated with partonic twist-three
and transversity processes, can be neglected.
We relate the first harmonics of the charge odd beam spin asymmetry and the charge asymmetry to
the imaginary part and real part of twist-two associated CFF ${\cal H}\approx {\cal H}_{++}$ by two linearized equations
\begin{eqnarray}
\label{A2CFF}
\Asin{1}{LU,I} \approx N \mbox{$c$}^{-1}_{\Im{\rm m}} \Im{\rm m}{\cal H}
\quad\mbox{and}\quad
\Acos{1}{C} \approx N \mbox{$c$}^{-1}_{\Re{\rm e}}  \Re{\rm e} {\cal H} \,,
\end{eqnarray}
where the coefficients are calculated from the theoretical expressions
\begin{eqnarray}
\label{c_^{-1}}
c_{\Im{\rm m}}^{-1} = \frac{\partial\Asin{1}{LU,I}}{\partial \Im{\rm m}{\cal H} }\Bigg|_{{\cal F}=0}
\quad\mbox{and}\quad
c_{\Re{\rm e}}^{-1} = \frac{\partial \Acos{1}{C}}{\partial \Re{\rm e} {\cal H} }\Bigg|_{{\cal F}=0}\,.
\end{eqnarray}
In this procedure, we set the DVCS-squared term in the denominator to zero which, however, appears in the normalization factor $N$.
To a good approximation, this overall factor can be also
expressed by the ratio of the BH and DVCS cross sections
\begin{eqnarray}
\label{eq:N}
0 \lesssim  N(\mbox{\boldmath $A$}) \approx \frac{\int_{-\pi}^\pi\! d\phi\,w(\phi)d\sigma_{\rm BH}(\phi)}{\int_{-\pi}^\pi\! d\phi\, w(\phi)\left[d\sigma_{\rm BH}(\phi)+ d\sigma_{\rm DVCS}(\phi)\right]} =\frac{1}{1+ \frac{k}{4} |\cffH(\mbox{\boldmath $A$})|^2 } \lesssim 1\,,
\end{eqnarray}
where $k$ is a known kinematical factor.
Since this overall factor depends on $|\cffH |$, it can be equivalently viewed as a function of
the asymmetries and of $N$. Plugging  the solution
\begin{eqnarray}
\label{As2CFFs-spin0-tw2}
\Im{\rm m} \cffH = \frac{\mbox{$c$}_{\Im{\rm m}} }{N(\mbox{\boldmath $A$})} \Asin{1}{LU,I}
\quad\mbox{and}\quad
\Re{\rm e} \cffH = \frac{\mbox{$c$}_{\Re{\rm e}}}{N(\mbox{\boldmath $A$})}\Acos{1}{C}\,,
\end{eqnarray}
into (\ref{eq:N}) yields a cubic equation in $N$ that has {\em two} non-trivial solutions:
\begin{eqnarray}
\label{N-spin0-tw2-sol}
N(\mbox{\boldmath $A$})\approx \frac{1}{2} \left(1 \pm \sqrt{1-k\,c_{\Im{\rm m}}^2 \left(\Asin{1}{LU,I}\right)^2 -k\, c_{\Re{\rm e}}^2\left(\Acos{1}{C}\right)^2 }\right).
\end{eqnarray}
In HERMES kinematics  the unpolarized BH cross section overwhelms the DVCS
one. Hence, we take the solution with the positive root which satisfies the
condition $N(\mbox{\boldmath $A$}=\mbox{\boldmath $0$})=1$ %
\footnote{The solution (\ref{N-spin0-tw2-sol}) with the negative root satisfies the boundary
condition $N(\mbox{\boldmath $A$}=\mbox{\boldmath $0$})=0$
and it is the one to take if the unpolarized DVCS cross section is larger than the BH one.}.
Finally,  for normally distributed random variables we can propagate the variances in the known manner rather than
discuss the map of probability distributions.

\begin{figure}[t]
\begin{center}
\includegraphics[scale=0.52]{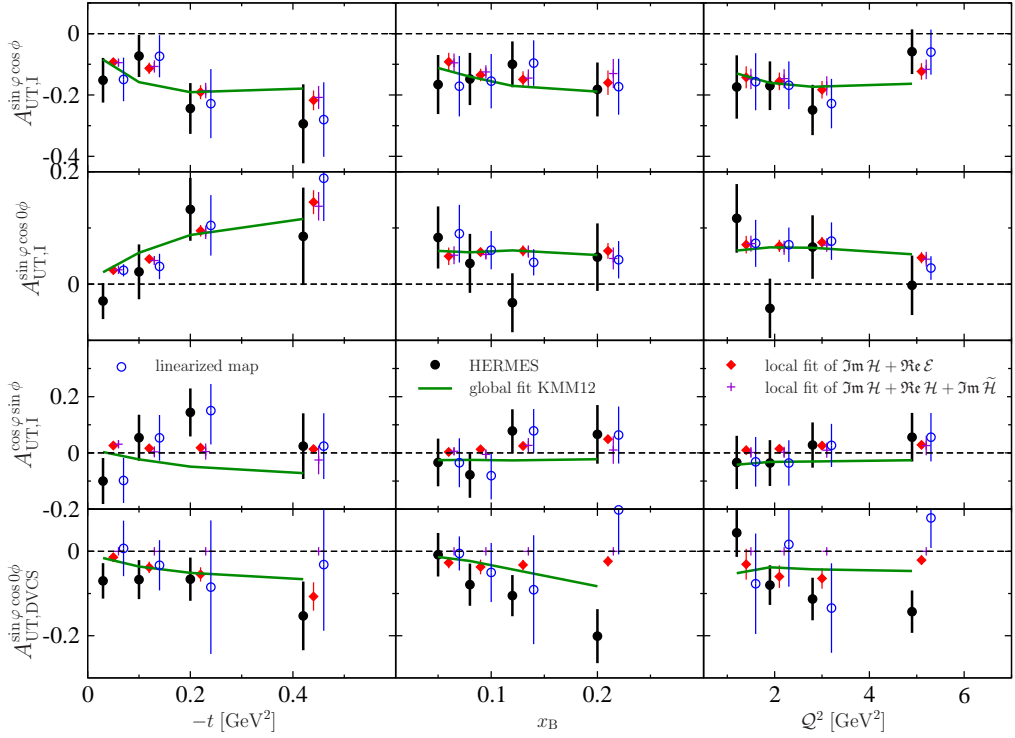}
\end{center}
\vspace{-4mm}
\caption{\small Fits  to harmonics of asymmetries of scattering on an \emph{unpolarized} target.
Black dots are H{\sc ermes} data with systematic errors added in quadrature. Local fits in two
different scenarios are shown as red diamonds (fit to $\Im{\rm m}\mathcal{H}$ and $\Re{\rm e}\mathcal{E}$) and
blue pluses (fit to $\Im{\rm m}\mathcal{H}$, $\Re{\rm e}\mathcal{H}$, and $\Im{\rm m}\widetilde{\mathcal{H}}$),
slightly displaced to the right for legibility.
For comparison, we also show the result of a global fit to world DVCS data as a green solid line.}
\label{fig:unp}
\vspace{-5mm}
\end{figure}
In our analyzes we employ only twist-two dominated  asymmetries from the final set of DVCS off-the-proton data from the
H{\sc ermes} collaboration extracted using a missing-mass event
selection method \cite{Airapetian:2008aa,Airapetian:2010ab,Airapetian:2011uq,Airapetian:2012mq}, which are used to extract the four twist-two associated CFFs $\{{\cal H},{\cal E},\widetilde{\cal H},\overline{\cal E}\approx\xB\widetilde{\cal E}/(2-\xB)\}$. To find the one-to-one map for the BH dominated scenario, we numerically solve eight quadratic equations for the following four single spin
and charge as well as double spin asymmetries,
\begin{eqnarray}
\label{Asin-HERMES}
\quad
\left(\!\!
\begin{array}{c}
\Asin{1}{LU,I} \\
\Asin{1}{UL,+}\\
\Asincos{1}{UT,I}\\
\Acossin{1}{UT,I} \\
\end{array}
\!\!\right)
\quad \Rightarrow \quad
\Im{\rm m}\!\left(\!\!
\begin{array}{c}
\cffH \\
\cfftH \\
\cffE \\
\cffbE \\
\end{array}
\!\!\right),
\qquad
\left(\!\!
\begin{array}{c}
\Acos{1}{C} \\
\Acos{1}{LL,+}\\
\Asinsin{1}{LT,I}\\
\Acoscos{1}{LT,I} \\
\end{array}
\!\!\right)
\quad \Rightarrow \quad
\Re{\rm e}\!\left(\!\!
\begin{array}{c}
\cffH \\
\cfftH \\
\cffE \\
\cffbE \\
\end{array}
\!\!\right).
\end{eqnarray}
The predictions from our one-to-one map for three charge odd $\cos(0\phi)$ harmonics $\Acos{1}{C}$, $\Asinsin{1}{LT,I}$, and $\Acoscos{1}{LT,I}$, which are correlated with the $\cos\phi$ harmonics, the charge even harmonics $\Asincos{0}{UT,DVCS}$, $\Acoscos{0}{LT,BH+DVCS}$, as well as the $\Acos{0}{LL,+}$ harmonic, which is dominated by the BH squared term, are consistent with the HERMES measurements.  In Fig.~\ref{fig:methods} we show the resulting CFFs from the one-to-one map (stars), a linearized map (circles),  and a least square fit (triangles) to all fourteen asymmetries.  The results are in general consistent, however, in  bin
\#3 and \#8 the fitting routine picked up the DVCS dominated solution rather the BH one. For \#3 we cured this by constraining  $\Re{\rm e} {\cal E}$,
which yields in return smaller error bars for other sub-CFFs. As one can see only $\Im{\rm m}{\cal H}$ significantly differs from zero while  $\Re{\rm e}{\cal H}$ and also $\Im{\rm m}\widetilde{\cal H}$ are compatible with zero and well constrained. All other  sub-CFFs are rather noisy and compatible with zero, too.

We also confronted our model ansatz from \cite{Kumericki:2009uq}, designed for the extraction
of the dominant CFF $\cal H$ from unpolarized proton DVCS data, with the world DVCS data set.
The resulting $\chi/{d.o.f.} \approx 1.6$ fit is strictly speaking not
a good fit, but it is acceptable for a global fit to data coming
from such a variety of experiments and observables. In particular tension is induced by the unpolarized HALL A cross section
measurement at four different $-t$ values \cite{Munoz_Camacho:2006hx} with a `big' $\widetilde{\cal H}$ scenario and longitudinally polarized proton spin asymmetry measurements. We also note that in our
model $\Im{\rm m} {\cal E}$ is set to zero, however, the transverse target HERMES data are well described, see Fig.~\ref{fig:unp}.

\section{Summary}
In the first decade of systematic measurements of exclusive processes at medium and high energies it has been shown that the GPD framework
can be utilized to describe DVCS and even DVMP data.  It is expected that
a global fit to all channels seems to be feasible within the collinear factorization approach in which unobserved
transverse degrees of freedom are integrated out. It also became obvious that GPD $H$ is dominant, while an phenomenological access to GPD $E$ cannot be
reached from present data. Such a goal requires high-luminosity experiments with dedicated detectors as planned at JLAB@12GeV experiments and
at a proposed Electron-Ion-Collider \cite{Deshpande:2012bu}.

\section*{Acknowledgments}

This work was partly supported by the Scottish
Universities Physics Alliance, the UK's Science and Technology
Facilities Council, by the Joint Research Activity
''Study of Strongly Interacting Matter'' (acronym HadronPhysics3, Grant Agreement No.~283286) under the Seventh Framework Program of the European Community,
and by Croatian Ministry of Science, Education and Sport, contract no.
119-0982930-1016.


\end{document}